\documentstyle[12pt]{article}

\def\Lc{{\cal L}}

\font\tenbb=msym10
\font\sevenbb=msym7
\font\fivebb=msym5
\newfam\bbfam
\textfont\bbfam=\tenbb \scriptfont\bbfam=\sevenbb
\scriptscriptfont\bbfam=\fivebb
\def\bb{\fam\bbfam}

\def\Rb{{\bb R}}

\def\build#1_#2^#3{\mathrel{
\mathop{\kern 0pt#1}\limits_{#2}^{#3}}}

\title{\huge Photon rockets and gravitational radiation}
\author{\Large Thibault Damour \\
\\
\normalsize Institut des Hautes Etudes Scientifiques \\
\normalsize 91440 Bures-sur-Yvette, France \\
\\
\small and \\
\\
\normalsize DARC, CNRS - Observatoire de Paris \\
\normalsize 92195 Meudon, France \\
\\}
\date{\small November 24, 1994}

\begin{document}

\maketitle

\bigskip

\begin{abstract}

The absence of gravitational radiation in Kinnersley's ``photon rocket''
solution of Einstein's equations is clarified by studying the mathematically
well-defined problem of point-like photon rockets in Minkowski space (i.e.
massive particles emitting null fluid anisotro\-pically and accelerating
because of the recoil). We explicitly compute the (uniquely defined) {\it
linearized} retarded gravitational waves emitted by such objects, which are the
coherent superposition of the gravitational waves generated by the motion of
the
massive point-like rocket and of those generated by the energy-momentum
distribution of the photon fluid. In the special case (corresponding to
Kinnersley's solution) where the anisotropy of the photon emission is purely
dipolar we find that the gravitational wave amplitude generated by the
energy-momentum of the photons exactly cancels the usual $1/r$ gravitational
wave amplitude generated by the accelerated motion of the rocket. More general
photon anisotropies would, however, generate genuine gravitational radiation at
infinity. Our explicit calculations show the compatibility between the
non-radiative character of Kinnersley's solution and the currently used
gravitational wave generation formalisms based on post-Minkowskian perturbation
theory.

\end{abstract}

\newpage

\section{Introduction}

In 1969 Kinnersley \cite{K69} constructed an exact solution of Einstein's
equations containing four arbitrary functions of time. In the literature (see
e.g. \cite{KSHM}) this solution is referred to as a ``photon rocket'' and is
interpreted as a ``particle emitting null fluid anisotropically, and
accelerating because of the recoil'' \cite{B94}. Recently, Bonnor \cite{B94}
pointed out a paradox of this interpretation: one would expect the presence of
gravitational radiation emitted by the ``accelerated particle'', while, as
first shown by Kinnersley \cite{K69}, and recently confirmed by Bonnor
\cite{B94}, the spacetime described by Kinnersley's solution is asymptotically
flat and (though non-stationary) contains no gravitational radiation at
infinity. Technically, if, say, the ``mass function'' $m(s)$ is constant for
$s\leq s_0$ and becomes smoothly variable for $s\geq s_0$, Kinnersley's
spacetime is identical to Schwarzschild's spacetime near past null infinity,
and has a (everywhere Petrov type D) Weyl tensor decreasing like $1/r^3$ (in
suitable, radiative, coordinates) near future null infinity, with a Ricci
tensor decreasing like $1/r^2$ corresponding to the energy-momentum of the
outgoing photon fluid (see equations (11) and (12) of \cite{K69}).

\medskip

One possible reaction to this paradox is to remark that the ``interpretation''
on which it is based cannot be given any rigourous meaning within the exact,
curved spacetime framework of Einstein's theory. Indeed, the notion of ``point
particle'' has no mathematical meaning in general relativity, and even if we
somehow ``fill in'' Kinnersley's solution to deal with a regular, extended
massive source, the notion of the ``position'' of the source (and of its
``quadrupole moment'') will have no unique definition, so that there will be no
unambiguous way of saying that some massive object is
accele\-rated and ``must'' therefore radiate at future null
infinity\footnote{Remembering that Kinnersley's solution can be taken to
coincide with a static Schwarzschild spacetime before some retarded time, we do
not have to worry about the possible presence of incoming gravitational
radiation.}. We believe that these ambiguities in the interpretation of
Kinnersley's solution are indeed very serious. However, we shall here tackle
directly the physics behind the paradox raised by Bonnor by noticing that there
is a mathematically well-defined si\-tuation in which we can self-consistently
describe a recoiling photon rocket and compute the retarded gravitational waves
it generates. More precisely, we consider, within the framework of {\it special
relativity}, a point-like massive particle (described by a $\delta$ function on
a
time-like worldline) emitting null fluid anisotropically. Working rigourously
within the framework of distribution theory \cite{S66}, we prove in section 3
(after having introduced our notation for various quantities associated with a
generic worldline in section 2) that the total energy-momentum tensor $T_{\mu
\nu} = T_{\mu \nu}^{({\rm mass \ point})} + T_{\mu \nu}^{({\rm photons})}$ is
everywhere conserved when the particle loses mass and recoils as expected from
the fluxes of photon energy and momentum at infinity. In section 4, we then
compute explicitly the (retarded) {\it linearized} gravitational field
generated
by the total $T_{\mu \nu}$. We find that the amount of gravitational radiation
at
infinity depends on the anisotropies of multipole order $\ell \geq 2$ in the
photon flux, there being a cancellation between the gravitational wave
amplitude
emitted by the energy-momentum distribution of the monopolar and dipolar photon
flux, and the gravitational wave amplitude emitted by the accelerated massive
particle. We conclude in section 5 that Bonnor's paradox\footnote{Remember that
a paradox is a ``statement that seems to say something opposite to common sense
or the truth; but which may contain a truth'' \cite{H74}.} contains some truth,
but that there is no incompatibility between the non-radiative character of
Kinnersley's solution (with its purely dipolar photon flux anisotropy) and the
consistency of the standard gravitational wave generation formalisms in
ge\-neral relativity, based on post-Minkowskian perturbation theory (such as a
recently developed multipolar-post-Minkowskian formalism \cite{BD86},
\cite{BD88}, \cite{BD89}, \cite{DI91a}, \cite{BD92}). Concerns recently raised
in the literature \cite{NS} about the reliabi\-lity of the current predictions
for the gravitational waves generated by, e.g., binary systems are therefore
totally unjustified.

\section{Geometry of Minkowski worldlines}

Let $\Lc$ denote a timelike worldline $z^{\mu} (s)$, parametrized by the proper
time $s$, in Minkowski space $(\Rb^4 ,\eta_{\mu \nu})$ with $\eta_{\mu \nu}
\equiv {\rm diag} (-1,+1,+1,+1)$. The four-velocity along $\Lc$ is denoted
$u^{\mu} (s) \equiv dz^{\mu} (s)/ds$ and satisfies $\eta_{\mu \nu} \ u^{\mu} \
u^{\nu} = -1$. To each point $x^{\mu}$ in Minkowski space one associates a
unique ``retarded'' point $z_R^{\mu} (x^{\nu})$ on $\Lc$, with parameter $s_R
(x^{\mu})$, defined as the intersection of $\Lc$ with the past light cone with
vertex at $x^{\mu}$. In equations, $z_R^{\mu} (x) \equiv z^{\mu} (s_R (x))$
with the function $s_R (x)$ defined as the only solution of
$$
\eta_{\mu \nu} \left( x^{\mu} -z^{\mu} \ [s_R (x)] \right) \ \left( x^{\nu}
-z^{\nu} \ [s_R (x)] \right) = 0 \eqno (2.1)
$$
such that $x^0 -z_R^0 > 0$. One also introduces the null vectors
$$
y^{\mu} (x) \equiv x^{\mu} - z_R^{\mu} (x) \ , \eqno (2.2{\rm a})
$$
$$
k^{\mu} (x) \equiv \frac{x^{\mu}-z_R^{\mu}(x)}{\rho_R (x)} \ , \eqno (2.2{\rm
b})
$$
where
$$
\rho_R (x) \equiv -(x^{\mu} -z_R^{\mu}) \ u_{\mu}^R \eqno (2.3)
$$
denotes the retarded distance between $x^{\mu}$ and the worldline $\Lc$. Here
and in the following the Minkowski metric is used to raise or lower indices.
Diffe\-rentiating the defining equation (2.1) yields the spacetime derivative
of
$s_R (x)$,
$$
\partial_{\mu} \ s_R (x) = - \frac{y_{\mu}(x)}{\rho_R (x)} = -k_{\mu} (x) \ ,
\eqno (2.4)
$$
from which follows
$$
\partial_{\mu} \ y^{\nu} = \delta_{\mu}^{\nu} +\rho_R^{-1} \ y_{\mu} \ u^{\nu}
=
\delta_{\mu}^{\nu} + k_{\mu} \ u^{\nu} \ , \eqno (2.5)
$$
$$
\partial_{\mu} \ \rho_R = \rho_R^{-1} \ y_{\mu} - u_{\mu} + \rho_R^{-1} \
(y\dot
u ) \ y_{\mu} = k_{\mu} -u_{\mu} +\rho_R (k \dot u ) \ k_{\mu} \ . \eqno (2.6)
$$
Here and in the following $\dot{u}^{\mu} \equiv du^{\mu} /ds$ denotes the
proper acceleration of the worldline $\Lc$ and we abbreviate Minkowski scalar
products as $(ab) \equiv a_{\mu} \ b^{\mu} \equiv \eta_{\mu \nu} \ a^{\mu} \
b^{\nu}$. It is also convenient to introduce notations for the unit spatial
vector representing the direction of $k^{\mu}$ in the 3-plane orthogonal to
$u^{\mu}$,
$$
k^{\mu} = u^{\mu} + n^{\mu} \ , \ {\rm with} \ (un) = 0 \ , \ (nn) = +1 \ ,
\eqno
(2.7)
$$
and for the projection operator orthogonal to $u^{\mu}$:
$$
\Delta_{\mu \nu} = \eta_{\mu \nu} + u_{\mu} \ u_{\nu} \ . \eqno (2.8)
$$
With this notation, and the results (2.4)-(2.6), we get also (suppressing the
``retarded'' label $R$ when there is no ambiguity)
$$
\partial_{\mu} \ \rho = n_{\mu} +\rho (n\dot u ) \ k_{\mu} \ , \eqno (2.9{\rm
a})
$$
$$
\partial_{\mu} \ k_{\nu} = \rho^{-1} (\Delta_{\mu \nu} -n_{\mu} \ n_{\nu}) -
(n\dot u ) \ k_{\mu} \ k_{\nu} \ , \eqno (2.9{\rm b})
$$
$$
\partial_{\mu} \ n_{\nu} = \rho^{-1} (\Delta_{\mu \nu} -n_{\mu} \ n_{\nu}) +
k_{\mu} \ \dot{u}_{\nu} -(n\dot u ) \ k_{\mu} \ k_{\nu} \ . \eqno (2.9{\rm c})
$$
As is well-known the null congruence $k^{\mu}$ is geodesic, shear-free and
expanding:
$$
k^{\nu} \ \partial_{\nu} \ k_{\mu} = 0 \ , \eqno (2.10{\rm a})
$$
$$
\partial_{\mu} \ k^{\mu} = -\Box \ s_R (x) = \frac{2}{\rho} \ , \eqno (2.10{\rm
b})
$$
$$
\partial^{\alpha} \ k_{\mu} \ \partial_{\alpha} \ k_{\nu} = \rho^{-2}
[\Delta_{\mu \nu} - n_{\mu} \ n_{\nu} ] \ , \eqno (2.10{\rm c})
$$
$$
\partial^{\mu} \ k^{\nu} \ \partial_{\mu} \ k_{\nu} = 2\rho^{-2} = \frac{1}{2}
\ (\partial_{\mu} \ k^{\mu})^2 \ . \eqno (2.10{\rm d})
$$
Finally, let us note that we can introduce ``retarded'' polar coordinates
centered on the worldline $\Lc$, say $(s,\rho ,\theta ,\varphi )$, such that
the polar coordinates of $x^{\mu}$ are $s_R (x)$, $\rho_R (x)$ and two angles
parametrizing the unit spatial vector $n^{\mu} (x)$, chosen, say, so as to
induce the standard metric $\rho^2 (d\theta^2 +{\rm sin}^2 \theta \ d\varphi^2
)$
on the 2-spheres $s=$ const, $\rho =$ const. This still leaves a large freedom
in
the rotation rate of the basis vectors $\partial n^{\mu} / \partial \theta$,
$({\rm sin} \theta)^{-1} \ \partial n^{\mu} /\partial \varphi$. We shall not
need
to restrict this freedom. One finds that the Minkowski volume element takes
(independently of the acceleration of the worldline) the familiar form
$$
d^4 x = \rho^2 \ ds \ d\rho \ d\Omega_{\bf n} \ , \eqno (2.11)
$$
with $d\Omega_{\bf n} = \sin \theta \ d\theta \ d\varphi$, while the surface
element of the retarded tubes $\rho =$ const. reads
$$
dS_{\mu} = \partial_{\mu} \ \rho_R \ \rho_R^2 \ ds \ d\Omega_{\bf n} \ . \eqno
(2.12)
$$

\section{Photon rockets in Minkowski space}

A physical photon rocket would be an extended massive object emitting (in the
WKB approximation) null fluid anisotropically. Let us prove that one can
consistently define, in Minkowski space, the point-like limit of such an
object. We consider an energy-momentum distribution which is partly
concentrated as a $\delta$-distribution on a time-like worldline $\Lc$, and
partly distributed as a null fluid spurting out of $\Lc$. Say
$$
T^{\mu \nu} (x) = T_{(m)}^{\mu \nu} (x) + T_{(p)}^{\mu \nu} (x) \ , \eqno (3.1)
$$
with a ``matter'' part
$$
T_{(m)}^{\mu \nu} (x) = \int ds \ m(s) \ u^{\mu} (s) \ u^{\nu} (s) \ \delta_4
(x^{\mu} - z^{\mu} (s)) \ , \eqno (3.2)
$$
and a ``photon'' part
$$
T_{(p)}^{\mu \nu} (x) = \frac{\varepsilon (s_R (x) , n^{\lambda} (x))}{4\pi \
\rho_R^2 (x)} \ k^{\mu} (x) \ k^{\nu} (x) \ . \eqno (3.3)
$$

\medskip

It has been verified by Kinnersley \cite{K69} and Bonnor \cite{B94} that the
energy-momentum tensor (3.3) entails an outgoing flux of energy and momentum at
infinity. They then argued heuristically that this must be compensated by a
loss of mass $dm(s)/ds$, and a recoil $du^{\mu} / ds$ of the ``central
particle''. Let us verify this more rigourously by studying whether the total
energy tensor $T^{\mu \nu} (x)$ can be {\it everywhere} conserved, including on
$\Lc$. This will prove the consistency of the point-like limit. We shall
consistently work in the framework of distribution theory, with $T^{\mu \nu}
(x)$ considered as a linear functional acting on smooth, compact-support
functions in $\Rb^4 : \langle T^{\mu \nu} ,\varphi \rangle = \int d^4 x \
T^{\mu \nu} (x) \ \varphi (x)$. For instance, $\langle T_{(m)}^{\mu \nu}
,\varphi \rangle = \int ds \ m(s) \ u^{\mu} \ u^{\nu} \ \varphi (z(s))$, while
the {\it distribution} $T_{(p)}^{\mu \nu} (x)$ is uniquely defined from the
{\it function} (3.3), in spite of its singular behaviour on the central
worldline $\Lc$, because $T_{(p)}^{\mu \nu}$ is locally integrable
[$O(\rho^{-2})$ singula\-rity only as $\rho \rightarrow 0$]. The distributional
derivative of the matter part (3.2) is easily verified to give
$$
\partial_{\nu} \ T_{(m)}^{\mu \nu} (x) = +\int ds \ \frac{d}{ds} \ (m(s) \
u^{\mu}) \ \delta_4 (x-z(s)) \ . \eqno (3.4)
$$
The computation of $\partial_{\nu} \ T_{(p)}^{\mu \nu}$ is more tricky. It
cannot be directly done on the function (3.3) because $\partial_{\nu} \
T_{(p)}^{\mu \nu} (x)$ has a non locally integrable singularity,
$O(\rho^{-3})$, along $\Lc$. One way to do it is to apply the distributional
definition, $\langle T_{(p)}^{\mu \nu} , - \partial_{\nu} \ \varphi \rangle$,
and use Stokes formula, with equation (2.12), when integrating by parts on a
domain $\rho_R (x) \geq \epsilon$ before letting $\epsilon \rightarrow 0$. A
more convenient technique, which is quite standard in distribution theory for
functions ha\-ving power-law singularities (see e.g. \cite{S66}), is to
introduce
a complex para\-meter $B$ and to work with the analytic continuation of
integrals
of ordinary functions. For instance, we can multiply equation (3.3) by $\rho^B$
and compute $\partial_{\nu} (\rho^B \ T_{(p)}^{\mu \nu} (x))$ which is a
locally integrable function for  $Re (B) > 0$. The looked for distribution is
then obtained by analytically continuing $B$ to zero. Using the derivatives
recalled in section 2, this gives after an easy calculation
$$
\partial_{\nu} (\rho^B \ T_{(p)}^{\mu \nu}) = B \ \rho^{B-3} \
\frac{\varepsilon (s_R ,n^{\lambda})}{4\pi} \ k^{\mu} \ . \eqno (3.5)
$$

\medskip

The problem is then reduced to evaluating the analytic continuation down to
$B=0$ of the four-dimensional integral
$$
I(B) = \int d^4 x \ \partial_{\nu} (\rho^B \ T_{(p)}^{\mu \nu}) \ \varphi (x) =
\frac{B}{4\pi} \int ds \ d\rho \ d\Omega_{\bf n} \ \rho^{B-1} \ \varepsilon
(s,n) \ k^{\mu} \ \varphi (x) \ , \eqno (3.6)
$$
where we have used retarded polar coordinates and equation (2.11). The $B$
factor in front shows that only an arbitrarily small neighbourhood of $\Lc$,
say $\rho \leq \rho_0$, matters. Using the Taylor expansion of $\varphi (x)$
gives $\varphi [x(s,\rho ,\theta ,\varphi )]= \varphi [z(s)] + O(\rho)$ when
$\rho \rightarrow 0$. As the analytic continuation of $B
\int_{0}^{\rho_0} d\rho \ \rho^{B-1} \ \rho = B \ \rho_0^{B+1} / (B+1)$ is
zero, we see that only the value of $\varphi$ along $\Lc$, $\varphi [z(s)]$,
contributes. The integration over the angles introduces the quantity
$$
\frac{1}{4\pi} \int d\Omega_{\bf n} \ \varepsilon (s,{\bf n}) \ k^{\mu}
=\varepsilon_0 (s) \ u^{\mu} + \varepsilon_1^{\mu} (s) \eqno (3.7)
$$
with
$$
\varepsilon_0 (s) \equiv \frac{1}{4\pi} \int d\Omega_{\bf n} \ \varepsilon
(s,{\bf n}) \ , \eqno (3.8)
$$
$$
\varepsilon_1^{\mu} (s) \equiv \frac{1}{4\pi} \int d\Omega_{\bf n} \ n^{\mu} \
\varepsilon (s,{\bf n}) \ , \eqno (3.9)
$$
while the integration over $\rho$ gives, modulo terms that tend to zero when
$B$ is analytically continued to zero, $B \int_{0}^{\rho_0} d\rho \
\rho^{B-1} = \rho_0^B = e^{B{\rm ln}\rho_0}$, which tends to one when $B$ is
continued down to zero, independently of $\rho_0$. Finally, this gives
$$
I(0) = \langle \partial_{\nu} \ T_{(p)}^{\mu \nu} ,\varphi \rangle = \int ds
(\varepsilon_0 (s) \ u^{\mu} + \varepsilon_1^{\mu} (s)) \ \varphi [z(s)] \ ,
\eqno (3.10)
$$
which means that we have proven the distributional result
$$
\partial_{\nu} \ T_{(p)}^{\mu \nu} (x) = \int ds (\varepsilon_0 (s) \ u^{\mu}
+\varepsilon_1^{\mu} (s)) \ \delta_4 (x-z(s)) \ . \eqno (3.11)
$$
Combining (3.11) with (3.4), we conclude that the total energy-momentum tensor
(3.1) will be everywhere conserved in the sense of distribution theory,
$\langle \partial_{\nu} \ T^{\mu \nu} ,\varphi \rangle = 0$, if and only if
$$
\frac{d}{ds} \ (m(s) \ u^{\mu} (s)) + \varepsilon_0 (s) \ u^{\mu}
+\varepsilon_1^{\mu} (s) =0 \ . \eqno (3.12)
$$
Noting that $u_{\mu} \ \varepsilon_1^{\mu} =0$ from its definition (3.9),
equation (3.12) is equivalent to
$$
\dot m =-\varepsilon_0 \ , \eqno (3.13{\rm a})
$$
$$
m \ \dot{u}^{\mu} = -\varepsilon_1^{\mu} \ . \eqno (3.13{\rm b})
$$
These are the results of \cite{B94} obtained here without relying on a
heuristic energy-momentum balance argument.

\section{Linearized gravitational waves emitted by photon rockets}

The conserved energy-momentum tensor (3.1) can be consistently taken as
distributional source term for linearized gravity. The linearized Einstein
equations for $g_{\mu \nu} = \eta_{\mu \nu} + h_{\mu \nu}$ (or the field
equations of the massless spin-2 field $h_{\mu \nu}$) read
$$
\Box \ h_{\mu \nu} + \partial_{\mu \nu} \ h - \partial_{\mu \alpha} \
h_{\nu}^{\alpha} - \partial_{\nu \alpha} \ h_{\mu}^{\alpha} = -16\pi \
G\left(T_{\mu \nu} -\frac{1}{2} \ \eta_{\mu \nu} \ T\right) \ , \eqno (4.1)
$$
where $\Box \equiv \eta^{\alpha \beta} \ \partial_{\alpha \beta}$, $h\equiv
h_{\alpha}^{\alpha}$, $T\equiv T_{\alpha}^{\alpha}$, indices being raised and
lowered by the Minkowski metric. To make contact with usual perturbation
techniques in general relativity let us start by looking for solutions of
equation (4.1) in harmonic gauge
$$
\partial^{\nu} \ h_{\mu \nu}^{({\rm har})} - \frac{1}{2} \ \partial_{\mu} \
h^{({\rm har})} = 0 \ . \eqno (4.2)
$$
This yields as usual
$$
\Box \ h_{\mu \nu}^{({\rm har})} = -16 \pi \ G\left(T_{\mu \nu} - \frac{1}{2} \
\eta_{\mu \nu} \ T \right) \ . \eqno (4.3)
$$
To simplify the discussion we assume that the photon rocket started
(smooth\-ly)
its activity only at some finite time in the past, i.e. $\varepsilon
(s,n^{\mu}) = 0$ for $s\leq s_0$. This implies that $m(s)$ was constant and
$\Lc$ straight before $s_0$ (see equations (3.13)), and that the non
compact-support part of $T^{\mu \nu} (x)$, namely $T_{(p)}^{\mu \nu} (x)$,
equation (3.3), has its support entirely contained in the forward light cone
with vertex $z^{\mu} (s_0)$. These properties ensure that the {\it retarded}
solution of (4.3), i.e. the convolution of the retarded Green's function,
$$
G_R (x^{\mu}) \equiv - \frac{1}{2\pi} \ \theta (x^0) \ \delta (\eta_{\mu \nu} \
x^{\mu} \ x^{\nu}) \equiv - \frac{1}{4\pi} \ \frac{\delta (x^0 -\vert {\bf x}
\vert)}{\vert {\bf x} \vert} \ ; \ \Box \ G_R (x) = \delta_4 (x) \eqno (4.4)
$$
with the right-hand side of (4.3), is mathematically well-defined\footnote{In
particular, it is given by a compact-support three-dimensional integral; the
``bad'' $\sim 1/r^2$ behaviour of the source term at future null infinity
causing no convergence problem in the retarded potential integral.}, and can be
characterized as being the only solution of (4.3) containing no incoming
radiation. We write it as
$$
h_{\mu \nu}^{({\rm har})} (x) = -16\pi \ G \ \Box_R^{-1} \left(T_{\mu \nu} -
\frac{1}{2} \ \eta_{\mu \nu} \ T\right) \ , \eqno (4.5)
$$
where $(\Box_R^{-1} \ S)(x) \equiv \int d^4 y \ G_R (x-y) \ S(y)$ denotes the
retarded potential operator.

\medskip

The linear split (3.1) of $T_{\mu \nu}$ leads to a corresponding linear split
of $h_{\mu \nu}^{({\rm har})}$, say
$$
h_{\mu \nu}^{({\rm har})} (x) = h_{\mu \nu}^{(m)} (x) + h_{\mu \nu}^{(p)} (x) \
. \eqno (4.6)
$$
The ``matter'' part $h_{\mu \nu}^{(m)}$ is obtained in a standard way from its
source (3.2) using the covariant form of $G_R (x^{\mu} -z^{\mu} (s)) \ \propto
\ \delta [(x_{\mu} -z_{\mu} (s))(x^{\mu} - z^{\mu} (s))]$ and the formula
$\delta [\varphi (s)] = \ \build \sum_{n}^{} \ \delta (s-s_n) / \vert \varphi'
(s_n)\vert$ where $n$ labels the roots $s_n$ of $\varphi (s) = 0$:
$$
h_{\mu \nu}^{(m)} (x) = 2G \left( \frac{m(s) [2u_{\mu} \ u_{\nu} +\eta_{\mu
\nu}]}{\rho} \right)_R \ . \eqno (4.7)
$$
Here, as in section 2, the suffix $R$ indicates that all $s$-dependent
quantities must be evaluated at the retarded point $s_R (x)$
(Li\'enard-Wiechert potential).

\medskip

The ``photon'' part $h_{\mu \nu}^{(p)}$ is
given by the following retarded potential integral
$$
h_{\mu \nu}^{(p)} (x) = -4G \ \Box_{R}^{-1} \ \left( \frac{\varepsilon (s_R ,n)
\ k_{\mu} \ k_{\nu}}{\rho_R^2} \right) \ , \eqno (4.8)
$$
which is well-defined as the integral of a function (rather than a
distribution) because of the local integrability of $1/\rho^2$. The integral
(4.8) can be analytically reduced to a one-dimensional integral over the proper
time $s$ (on the domain $s\leq s_R (x)$) by using the formula (2.24) of Ref.
\cite{BD92}\footnote{To generalize the result of \cite{BD92} which assumes a
straight worldline $\Lc$ to the present case of a curved worldline, one needs
(beyond multipole-expanding $\varepsilon (n)$) to represent the  source term in
(4.8) as a one-dimensional integral of elementary contributions with support on
light cones with vertex on $\Lc$.}. However, a more convenient way of studying
$h_{\mu \nu}^{(p)}$ is, following the study of the $u^{\alpha \beta}$ term in
Ref. \cite{BD92}, which has the same structure as $h_{\mu \nu}^{(p)}$ (for a
straight worldline), to introduce a simplifying gauge transformation. The role
of this gauge transformation (or li\-nearized coordinate transformation) is to
transform the source $\propto \ k_{\mu} \ k_{\nu} /\rho^2$ in equation (4.8)
into
a source proportional to $1/\rho^3$ which turns out to be much simpler to deal
with\footnote{Another utility of that gauge transformation is to deal away with
the terms of order $({\rm ln} \ \rho)/\rho$ in $h_{\mu \nu}^{({\rm har})}$ at
future null infinity. As is well-known these terms, due to the slow $1/\rho^2$
fall off of $T_{\mu \nu}$, arise in harmonic coordinates but can be gauged away
by introducing some suitable, ```radiative'' coordinates.}. However, in so
doing
we have to tackle non locally integrable terms $\sim \rho^{-3}$. This situation
is technically very similar to what happened in section 3 where we needed to
evaluate the distributional derivative of $T_{(p)}^{\mu \nu} \sim \rho^{-2}$,
i.e. to make sense of integrals involving $\partial_{\lambda} \ T_{(p)}^{\mu
\nu} \sim \rho^{-3}$. Like in section 3, a technically convenient, and
mathema\-tically rigourous, way of dealing with such terms is to introduce a
complex parameter $B$ and to work with the analytic continuation of certain
integrals.

\medskip

Given a function $\sigma (s,n^{\mu})$ which reduces to an angle-independent
constant $\sigma_0$ for $s\leq s_0$ (i.e. when $\varepsilon (s,n^{\mu})$
vanishes) and which will be related to $m(s)$ and $\varepsilon (s,n^{\mu})$
below, we define the gauge transformation $\xi_{\mu} (x)$ as follows. We first
consider the retarded potential integral
$$
\xi_{\mu}^B (x) \equiv \Box_R^{-1} \left( \frac{1}{2} \ \rho_R^{B-2} \ \sigma
(s_R (x) , n) \ k_{\mu} \right) \ , \eqno (4.9{\rm a})
$$
which depends on the complex parameter $B$, and is convergent when $-1 <
Re (B) <0$. Then we analytically continue $\xi_{\mu}^B (x)$ in $B$.
Generalizing standard arguments (see, e.g., Refs \cite{BD86}, \cite{BD88})
$\xi_{\mu}^B (x)$ is easily seen to admit a continuation as a meromorphic
function over the whole complex $B$ plane. Denoting by $FP_{B=0}$ (``Finite
Part
at $B=0$'') the operation of taking the constant term (zeroth power of $B$) in
the Laurent expansion of $\xi_{\mu}^B (x)$ around $B=0$, we then define
$$
\xi_{\mu} (x) \equiv FP_{B=0} \ (\xi_{\mu}^B (x)) \ . \eqno (4.9{\rm b})
$$
As $\rho^{-2}$ is locally integrable the analytic continuation factor $\rho^B$
is not needed in (4.9a) to deal with the neighbourhood of $\Lc$, but is useful
to deal with the slow fall-off $(\sim \rho^{-2})$ at past null infinity due to
the fact that $\sigma$ becomes a non-zero constant in the past. The retarded
integral (4.9a) is convergent when $-1 < Re(B) < 0$, and is easily seen to have
a simple pole at $B=0$ when analytically continued upwards in $Re (B)$. The
effect of this pole is to introduce a term proportional to $\sigma_0 \
u_{\mu}^0 \ \rho_0^B / B$ in $\xi_{\mu}^B$, and therefore, using the Laurent
expansion, $B^{-1} \ \rho_0^B = B^{-1} \ \exp (B \ {\rm ln} \ \rho_0) = B^{-1}
+{\rm ln} \ \rho_0 + O(B)$, a term proportional to $\sigma_0 \ u_{\mu}^0 \ {\rm
ln} \ \rho_0$ in $\xi_{\mu}$. [Here, $\rho_0$ denotes the spatial distance
between the field point $x^{\mu}$ and the incoming straight worldline with
four-velocity $u_{\mu}^0$]. This logarithmic term disappears when taking
$x$-derivatives, which corresponds to the fact that the $B$-dependent
integrals considered below will have a better convergence at past null
infinity, and generate no poles at $B=0$.

\medskip

Thanks to the properties of analytically continued integrals we can commute
$x$-derivatives with the integral operator $\Box_R^{-1}$. [When dealing with an
integral such as (4.9a), the integrand of which has power-law singularities
both on $\Lc$ and at past null infinity, it is convenient to split the spatial
domain of integration in two pieces, one of them being compact and centered on
$\Lc$; see, e.g., section 3 of \cite{BD88} for a similar situation.] This
yields
$$
\partial_{\mu} \ \xi_{\nu}^B = \frac{1}{2} \ \Box_R^{-1} \ \left[
\partial_{\mu} (\rho^{B-2} \ \sigma  \ k_{\nu}) \right] \ , \eqno (4.10)
$$
where, using the results of section 2,
$$
\partial_{\mu} (\rho_R^{B-2} \ \sigma (s_R ,n^{\alpha}) \ k_{\nu}) =
-\rho_R^{B-2} \ D\sigma \ k_{\mu} \ k_{\nu}
$$
$$
+ \rho_R^{B-3} \ \Bigl[\sigma \{ \Delta_{\mu \nu} + (B-2) \ n_{\mu} \ u_{\nu} +
(B-3) \ n_{\mu} \ n_{\nu} \}
$$
$$
+\frac{\partial \sigma}{\partial n_{\alpha}} \ (\Delta_{\alpha \mu} -n_{\alpha}
\ n_{\mu} ) \ k_{\nu}\Bigl] \ . \eqno (4.11)
$$
The quantity $D\sigma$ in (4.11) denotes the action of a certain first-order
diffe\-rential operator on $\sigma (s,n^{\alpha})$:
$$
D\sigma (s,n^{\alpha}) = \frac{\partial \sigma}{\partial s} + [(n\dot u ) \
k_{\alpha} - \dot{u}_{\alpha} ] \ \frac{\partial \sigma}{\partial n_{\alpha}} +
(3-B) (n \dot u ) \sigma \ . \eqno (4.12)
$$
Here we consider $\sigma (s,n^{\alpha})$ as some explicit function of $s$ and
the four components of $n^{\alpha}$, e.g. the one obtained by expanding the
angular dependence of $\sigma$ in cartesian symmetric-trace-free (STF)
polynomials in $n^{\alpha}$ [equivalent to a spherical harmonic expansion in
$Y_{\ell}^m (\theta , \varphi)$]
$$
\sigma (s,n^{\alpha}) = \sum_{\ell = 0}^{\infty} \sigma_{\alpha_1 \alpha_2
\ldots \alpha_{\ell}} \ (s) \ \widehat{n}^{\alpha_1 \alpha_2 \ldots
\alpha_{\ell}} \ , \eqno (4.13)
$$
where $\widehat{n}^{\alpha_1 \ldots \alpha_{\ell}}$ denotes the STF projection
(within the spatial hyperplane orthogonal to $u^{\alpha}$) of $n^{\alpha_1}
n^{\alpha_2} \ldots n^{\alpha_{\ell}}$. For instance, $\widehat{n}^{\mu \nu}
\equiv n^{\mu} \ n^{\nu} -\frac{1}{3} \ \Delta^{\mu \nu}$. Without loss of
generality, we can also consider that the $s$-dependent coefficients in (4.13)
are purely spatial $(u^{\alpha_1} \ \sigma_{\alpha_1 \ldots \alpha_{\ell}} =
\ldots = u^{\alpha_{\ell}} \ \sigma_{\alpha_1 \ldots \alpha_{\ell}} = 0)$ and
STF, in which case we can replace $\widehat{n}^{\alpha_1 \ldots \alpha_{\ell}}$
by $n^{\alpha_1} \ldots n^{\alpha_{\ell}}$ in (4.13) so that the $s$-derivative
in (4.12) acts only on $\sigma_{\alpha_1 \ldots \alpha_{\ell}} \ (s)$.
Evidently,
the quantities $D\sigma$ and $\partial \sigma / \partial n_{\alpha}
(\Delta_{\alpha \mu} - n_{\alpha} \ n_{\mu})$ entering (4.11) are defined
independently of the multipole expansion (4.13), but an alternative definition
would imply choosing some explicit parametrization of $n^{\alpha}$ in terms of
two polar angles $\theta , \varphi$ (i.e. choosing a triad of vectors
orthogonal
to $u^{\alpha}$) which we prefer to shun doing.

\medskip

We see from equation (4.11) that the gauge transformation $h'_{\mu \nu} =
h_{\mu \nu} + \partial_{\mu} \ \xi_{\nu} + \partial_{\nu} \ \xi_{\mu}$
generates terms proportional to $k_{\mu} \ k_{\nu} / \rho^2$ which can cancel
the original ``photon'' piece (4.8) and replace it by the retarded potential of
source terms $\propto \rho^{-3}$. To perform this transformation explicitly,
it is convenient to consider that the photon flux function $\varepsilon
(s,n^{\alpha})$ is, similarly to equation (4.13), decomposed in angular
multipoles
$$
\varepsilon (s,n^{\alpha}) = \sum_{\ell =0}^{\infty} \varepsilon_{\ell}
(s,n^{\alpha}) \ , \eqno (4.14)
$$
where $\varepsilon_{\ell} (n^{\alpha})$ is a linear combination of
$\widehat{n}^{\alpha_1 \ldots \alpha_{\ell}}$. The first two terms in the
multipole expansion (4.14), namely the monopole $\varepsilon_0$ and dipole
parts $\varepsilon_1$, are fully determined by the integrals (3.8) and (3.9).
More precisely $\varepsilon_0$ (defined by equation (4.14) as the monopole
piece of $\varepsilon (s,n)$) is identical with $\varepsilon_0 (s)$ defined by
equation (3.8), while the dipole piece of $\varepsilon (s,n)$ reads
$$
\varepsilon_1 (s,n^{\alpha}) = 3 \ \varepsilon_1^{\mu} (s) \ n_{\mu} \eqno
(4.15)
$$
in terms of the definition (3.9).

\medskip

Let us first consider the special case (that we shall the ``Kinnersley case'')
where the photon flux $\varepsilon (s,n)$ contains only a monopole and a dipole
piece:
$$
\varepsilon^K (s,n^{\alpha}) = \varepsilon_0 (s) + 3 \ \varepsilon_1^{\alpha}
(s) \ n_{\alpha} \ . \eqno (4.16)
$$
In view of the balance equations (3.13), this can be rewritten in terms of
``mechanical'' quantities
$$
\varepsilon^K (s,n^{\alpha}) = - [ \dot m (s) + 3 \ m(s) \ (\dot u \ n)] \ .
\eqno (4.17)
$$
Let us correspondingly consider the special case where $\sigma$ is purely
monopolar: $\sigma (s,n) = \sigma^K (s)$. From equations (4.10)-(4.12) (which
simplify when $\sigma$ depends only on $s$) we find for the gauge
transformation
$$
\partial_{\mu} \ \xi_{\nu}^{B(K)} + \partial_{\nu} \ \xi_{\mu}^{B(K)} =
\Box_R^{-1} \biggl\{ -\rho^{B-2} \ [ \dot{\sigma}^K + (3-B) \ \sigma^K \ (n\dot
u
)] \ k_{\mu} \ k_{\nu}
$$
$$
+\rho^{B-3} \ \sigma^K \left[ \frac{B}{3} \ \Delta_{\mu \nu} + \frac{B-2}{2} \
(n_{\mu} \ u_{\nu} + n_{\nu} \ u_{\mu}) + (B-3) \ \widehat{n}_{\mu \nu} \right]
\biggl\} \ , \eqno (4.18)
$$
where we have decomposed the functions of $n^{\alpha}$ in irreducible
multipoles of order $0,1$ and $2$ ($\widehat{n}^{\mu \nu} = n^{\mu} \ n^{\nu} -
\frac{1}{3} \ \Delta^{\mu \nu}$).

\medskip

Comparing (4.18) with the result of inserting
equation (4.17) into equation (4.8) we are led to defining
$$
\sigma^K (s) \equiv +4 \ G \ m(s) \ . \eqno (4.19)
$$
For this choice we see easily that the first term on the right-hand side of
(4.18) is analytic at $B=0$ (no pole; thanks to the vanishing of
$\dot{\sigma}^K$ and $\sigma^K (n\dot u )$ in the past) and that its value at
$B=0$ is
$$
+ \ 4G \ \Box_R^{-1} (\rho^{-2} \ \varepsilon^K (s_R ,n) \ k_{\mu} \ k_{\nu})
$$
which is precisely the {\it opposite} of $h_{\mu \nu}^{(p)} (x)$, defined in
equation (4.8), in the Kinnersley case.

\medskip

We are then left with evaluating three
types of terms: ``monopolar'' terms of the form $\Box_R^{-1} (B \ \rho^{B-3} \
\varphi (s))$, and ``dipolar'' and ``quadrupolar'' terms of the form
$\Box_R^{-1}
(\rho^{B-3} \ \varphi (s) \ \widehat{n}_{\alpha_1 \ldots \alpha_{\ell}})$ with
$\ell = 1$ or $2$. The monopolar term
$$
\Box_R^{-1} \left( \frac{B}{3} \ \rho_R^{B-3} \ \sigma^K (s_R) \ \Delta_{\mu
\nu} (s_R) \right) \eqno (4.20)
$$
appearing in the second square bracket on the right-hand side of (4.18) is
easily evaluated, when $B$ is analytically continued to zero, by the same
reasoning used in section 3 above to deal with the integral of the product of
the right-hand side of equation (3.5) with a smooth function [here that smooth
function is $G_R (x-x')$]. Denoting $AC_{B=0}$ the operation of taking
the analytic continuation at $B=0$, we deduce from the results of section 3
$$
AC_{B=0} \ \{ B \ \rho^{B-3} \ \varphi (s) \} = 4\pi \int ds \ \varphi (s) \
\delta_4 (x-z(s)) \ , \eqno (4.21)
$$
in the sense of distribution theory [where the $B$ factor in front ensures the
absence of pole at $B=0$]. Convoluting the Green's function (4.4) with (4.21),
we find that the analytic continuation of the monopolar term (4.20) yields
$$
-\frac{1}{3} \ \frac{\sigma^K (s_R) \ \Delta_{\mu \nu} (s_R)}{\rho_R} =
-\frac{4G}{3} \ \left( \frac{m(s) \ \Delta_{\mu \nu}}{\rho} \right)_R \ . \eqno
(4.22)
$$

\medskip

The dipolar and quadrupolar terms (last two terms on the right-hand side of
equation (4.18)) can be evaluated by using explicit formulas derived in Ref.
\cite{BD88}. There, indeed, it was shown that, in the case of a straight
central worldline $\Lc_0$, the analytic continuation at $B=0$ of the retarded
potential generated by $r^{B-k} \ H(t-r) \ \widehat{n}_{i_1 \ldots i_{\ell}}$
was expressible, when $3\leq k\leq \ell +2$, as a finite sum of derivatives of
$H(t-r)$ divided by powers of $r$. In particular, when $k=3$, $\ell \geq 1$ and
$H(t)$ is the one-dimensional delta function $\delta (t-\sigma)$, equation
(4.24) of Ref. \cite{BD88} reads
$$
AC_{B=0} \left\{ \Box_R^{-1} (r^{B-3} \ \delta (t-r-\sigma ) \ \widehat{n}_{i_1
\ldots i_{\ell}}) \right\} =
$$
$$
-\frac{1}{\ell (\ell +1)} \ \frac{\delta (t-r-\sigma)}{r}
\ \widehat{n}_{i_1 \ldots i_{\ell}} \ ; \ (\ell \geq 1) \ . \eqno (4.23)
$$
Here, the central straight worldline $\Lc_0$ is taken as time axis, $r$ denotes
the spatial distance between the field point $(x^{\mu})=(t,x^i)$ and $\Lc_0$,
$n^i \equiv x^i /r$, and $\sigma$ is a constant, parametrizing a specific point
$(\sigma,{\bf 0})$ on $\Lc_0$.

\medskip

Equation (4.23) can be generalized to the case, of interest for the present
paper, of a generic {\it curved} worldline $\Lc$ by noticing that if $\Lc_0$ is
taken to be the tangent to $\Lc$ at $z^{\mu} (\sigma)$ then $[\delta
(t-r-\sigma)]_{\Lc_0} = [\delta (s_R (x)-\sigma)]_{\Lc}$ as distributions in
$\Rb^4$. [Indeed, both distributions have the future light cone of $z^{\mu}
(\sigma)$ as support, and using the fact that the volume element in retarded
coordinates (2.11) does not depend on the acceleration $\dot{u}^{\mu}$ one sees
immediately that $\int d^4 x \ \delta (t-r-\sigma) \ \varphi (x) = \int d^4 x \
\delta (s_R (x)-\sigma) \ \varphi (x)$]. Having generalized (4.23) to an
elementary source with support on a light cone centered on $\Lc$, we can pass
to a general retarded-time-dependence of the source by integrating over the
parameter $\sigma$ using the decomposition $H(s_R (x)) = \int d\sigma \
H(\sigma) \ \delta (s_R (x)-\sigma)$. This yields
$$
AC_{B=0} \left\{ \Box_R^{-1} \left( \rho_R^{B-3} \ H(s_R (x)) \
\widehat{n}_{\mu_1 \ldots \mu_{\ell}} \right) \right\} =
$$
$$
-\frac{1}{\ell (\ell +1)} \ \frac{H(s_R (x))}{\rho_R} \ \widehat{n}_{\mu_1
\ldots
\mu_{\ell}} \ ; \ (\ell \geq 1) \ . \eqno (4.24)
$$
It is remarkable that, though the left-hand side, evaluated at the field point
$x^{\mu}$, is an integral over the past light cone of $x^{\mu}$ whose integrand
depends on the full past history of the source, i.e. on the value of the
function $H(s)$ over the interval $-\infty \leq s \leq s_R (x)$, the final
result on the right-hand side depends only on the value of $H(s)$ at the
retarded point $s_R (x)$. This simplification does not hold for integrands
proportional to $\rho_R^{-2}$, as they appeared in the original expression
(4.8)
before introducing the gauge transformation (4.9).

\medskip

Using the result (4.24) we can read off equation (4.18) the contributions of
the dipolar and quadrupolar terms in the last bracket. Namely:
$$
\frac{\sigma^K}{\rho_R} \ \left[ +\frac{1}{2} \ (n_{\mu} \ u_{\nu} + n_{\nu} \
u_{\mu}) + \frac{1}{2} \ \widehat{n}_{\mu \nu} \right] =
$$
$$
2G \ \frac{m(s_R)}{\rho_R} \ \left[ n_{\mu} \ u_{\nu} + n_{\nu} \ u_{\mu}
+\widehat{n}_{\mu \nu} \right] \ . \eqno (4.25)
$$
Gathering our results, we conclude that the gauge-transform of the original
harmonic linearized field (4.6),
$$
h'_{\mu \nu} (x) = h_{\mu \nu}^{({\rm har})} (x) + \partial_{\mu} \ \xi_{\nu} +
\partial_{\nu} \ \xi_{\mu} = h_{\mu \nu}^{(m)} (x) + h_{\mu \nu}^{(p)'} (x) \ ,
\eqno (4.26)
$$
with
$$
h_{\mu \nu}^{(p)'} (x) \equiv h_{\mu \nu}^{(p)} (x) + \partial_{\mu} \
\xi_{\nu} + \partial_{\nu} \ \xi_{\mu} \ , \eqno (4.27)
$$
is given, in the Kinnersley case (4.16), (4.17), by adding the matter
contribution (4.7) and the gauge-transformed photon one obtained by adding
(4.22) and (4.25):
$$
\left[ h_{\mu \nu}^{(p)'}\right]^K = 2G \ \frac{m(s_R)}{\rho_R} \ \left[
-\frac{2}{3} \ \Delta_{\mu \nu} + n_{\mu} \ u_{\nu} + n_{\nu} \ u_{\mu}
+\widehat{n}_{\mu \nu} \right]_R \ . \eqno (4.28)
$$
Replacing the definitions $\widehat{n}_{\mu \nu} \equiv n_{\mu} \ n_{\nu}
-\frac{1}{3} \ \Delta_{\mu \nu}$, $\Delta_{\mu \nu} = \eta_{\mu \nu} + u_{\mu}
\ u_{\nu}$ and eliminating $n_{\mu}$ in favour of $k_{\mu} \equiv u_{\mu} +
n_{\mu}$, the result (4.28) can be rewritten as
$$
\left[ h_{\mu \nu}^{(p)'} \right]^K = 2G \ \frac{m(s_R)}{\rho_R} \ \left[
-\eta_{\mu \nu} -2 \ u_{\mu} \ u_{\nu} + k_{\mu} \ k_{\nu} \right]_R \ . \eqno
(4.29)
$$
We recognize in the first two terms of the (gauge-transformed) photon
contribution (4.29) the {\it opposite} of the matter contribution (4.7). We
have therefore proven by an explicit computation that, after a suitable gauge
transformation (or linearized coordinate transformation) (4.9), there were
terms in the gravitational field generated by the photon energy-momentum tensor
which cancelled the usual, direct matter terms (4.7) to leave as final combined
gravitational field in the special Kinnersley case (4.16), (4.17) the net
result
$$
\left[ h'_{\mu \nu} (x) \right]^K = \left[h_{\mu \nu}^{({\rm har})}\right]^K +
\partial_{\mu} \ \xi_{\nu}^K + \partial_{\nu} \ \xi_{\mu}^K = 2G \
\frac{m(s_R)}{\rho_R} \ k_{\mu} \ k_{\nu} \ . \eqno (4.30)
$$
Because of the specific algebraic structure $\propto \ k_{\mu} \ k_{\nu}$ of
the last right-hand side of equation (4.30), the gravitational field $h_{\mu
\nu}^{'K}$ is easily verified not to contain any physical $1/r$ gravitational
waves at (future null) infinity. This can be seen either by taking the usual
transverse-traceless (TT) projection of the spatial components in some fixed
reference frame [using $k_i = n_i +O(1/r)$], or by checking, from the
derivatives given in section 2, that, at the $1/\rho$ level, $\partial_{\alpha
\beta} \ h_{\mu \nu}^{'K} \ \propto \ k_{\alpha} \ k_{\beta} \ k_{\mu} \
k_{\nu}$
which implies the vanishing of the linearized Riemann tensor at the $1/\rho$
level. [A more complete calculation, first performed by Kinnersley \cite{K69},
shows that the linearized Weyl tensor of (4.30) is of order $\rho^{-3}$ at
infinity, while its linearized Ricci tensor, proportional to $T_{\mu \nu}$, is
of order $\rho^{-2}$].

\medskip

The non radiative result (4.30) has been derived in the special case where the
anisotropy of the photon flux $\varepsilon (s,n)$ is purely dipolar, equations
(4.16), (4.17). Let us briefly discuss the generic case where $\varepsilon
(s,n)$ contains more general anisotropies. Using the notation (4.14), we can
always write
$$
\varepsilon (s,n) = \varepsilon^K (s,n) + \sum_{\ell =2}^{\infty} \
\varepsilon_{\ell} (s,n) \ , \eqno (4.31)
$$
keeping unchanged the definition (4.16), and the result (4.17) which followed
from the general balance equations (3.13). Let us then define $\sigma (s,n)$ as
the unique solution of the differential equation
$$
D_0 \ \sigma (s,n) = -4 \ G \ \varepsilon (s,n) \eqno (4.32)
$$
which reduces to the constant $4 \ G \ m(s_0)$ when $s\leq s_0$ (i.e. before
$\varepsilon (s,n)$ turns on). Here $D_0$ denotes the value at $B=0$ of the
differential operator (4.12). The (local) existence and uniqueness of $\sigma$
follow from the fact that, when written explicitely in terms of $s$ and two
angular coordinates $(\theta_A) = (\theta ,\varphi)$ on the sphere, $D_0 =
\partial / \partial s + v^A (s,\theta_B) \ \partial /\partial \theta_A$ is a
first-order linear operator. Given initial data for $\Lc$ in the past, and the
monopole and dipole parts, $\varepsilon_0$ and $\varepsilon_1$, for all values
of $s$ one can solve uniquely for the time evolution of $\Lc$ (via equations
(3.13)) and then for $\sigma (s,n)$. The linearity of equation (4.32) means
that its solution reads $\sigma = \sigma^K + \sigma^{\rm rad}$, where
$\sigma^K$
is the ``Kinnersley'' piece (4.19) and where $\sigma^{\rm rad}$ denotes the
solution of $D_0 \ \sigma^{\rm rad} = -4 \ G \ \build \sum_{\ell \geq 2}^{} \
\varepsilon_{\ell}$. From the explicit expression (4.12), one checks that, when
expanding $\sigma^{\rm rad}$ in orbital multipoles $\sigma_{\ell}^{\rm rad}
\propto \widehat{n}_{\alpha_1 \ldots \alpha_{\ell}}$, the $\ell = 0$ and $\ell
=1$ pieces of $D_0 \ \sigma^{\rm rad}$ are entirely expressible in terms of
$\sigma_0^{\rm rad}$ and $\sigma_1^{\rm rad}$ (without
the presence of a term $\propto \ \sigma_2^{\mu \nu} \ \dot{u}_{\mu} \
n_{\nu}$). As the source term for $D_0 \ \sigma^{\rm rad}$ has only multipoles
of order $\geq 2$, we conclude that $\sigma_0^{\rm rad} = \sigma_1^{\rm rad}
\equiv 0$, i.e.
$$
\sigma (s,n) = 4 \ G \ m(s) + \sum_{\ell =2}^{\infty} \sigma_{\ell} (s,n) \ ,
\eqno (4.33)
$$
where $\sigma_2 +\sigma_3 +\cdots$ is driven by $\varepsilon_2 +\varepsilon_3
+\cdots$.

\medskip

Using as above $\sigma$ to define the gauge transformation (4.9), we find from
equation (4.11) that the total gauge-transformed gravitational field reads
$$
h'_{\mu \nu} = h_{\mu \nu}^{({\rm har})} + \partial_{\mu} \ \xi_{\nu} +
\partial_{\nu} \ \xi_{\mu} = \left[ h'_{\mu \nu} \right]^K + h_{\mu
\nu}^{(p){\rm rad}} \ , \eqno (4.34)
$$
where $[h'_{\mu \nu}]^K$ is the non-radiative field (4.30) and

\newpage

$$
h_{\mu \nu}^{(p){\rm rad}} =
$$
$$
AC_{B=0} \biggl\{ \Box_R^{-1} \biggl( \rho_R^{B-3}
\biggl\{ \sigma^{\rm rad} \left[ \frac{B}{3} \ \Delta_{\mu \nu} + \frac{B-2}{2}
\ (n_{\mu} \ u_{\nu} + n_{\nu} \ u_{\mu}) + (B-3) \ \widehat{n}_{\mu \nu}
\right]
$$
$$
+\frac{1}{2} \ \frac{\partial \sigma^{\rm rad}}{\partial n_{\alpha}} \ \left[
(\Delta_{\alpha \mu} - n_{\alpha} \ n_{\mu} ) \ k_{\nu} + (\Delta_{\alpha \nu}
- n_{\alpha} \ n_{\nu} ) \ k_{\mu} \right] \biggl\} \biggl) \biggl\} \ . \eqno
(4.35)
$$
{}From formula (4.24), which can be compactly stated as
$$
AC_{B=0} \left\{ \Box_R^{-1} \ \left[ \rho_R^{B-3} \ H (s_R (x),n)\right]
\right\} = \rho_R^{-1} \ \Delta_n^{-1} \ H(s_R (x),n) \ , \eqno (4.36)
$$
where $\Delta_n$ is the Laplacian on the unit sphere, and $\Delta_n^{-1}$ its
inverse acting on zero-mean functions on the sphere, we get $h_{\mu
\nu}^{(p){\rm rad}} = \rho_R^{-1} \ U_{\mu \nu} (s_R ,n)$ where $U_{\mu \nu}
(s,n)$ is some explicit functional of $\sigma^{\rm rad} (s,n)$, which depends
only on the values of $\sigma^{\rm rad}$ for the same value of the proper time
$s$. This computation has been done explicitly in the case of a straight
worldline $\Lc$ in Ref. \cite{BD92}, see equations (2.20) there. In particular,
the TT projection of equation (2.20c) there [in which only the last term
$\propto (n_{L-2} \ \Pi_{ijL-2})^{\rm TT}$ survives] show that the $1/r$
gravitational wave amplitude radiated at infinity is non zero as soon as
$\sigma^{\rm rad} (n)$ contains a non-zero multipole of order $\ell \geq 2$.

\medskip

The conclusion is that, as soon as the photon flux function $\varepsilon (s,n)$
contains an anisotropy of multipole order $\ell \geq 2$, it will drive such
terms in $\sigma^{\rm rad} (s,n)$ which will, through equations (4.35), (4.36),
radiate genuine $1/r$ gravitational waves at future null infinity\footnote{Note
that in the limiting case of an infinitely massive photon rocket, $\Lc$ becomes
straight and $\sigma^{\rm rad} (s,n) = -4 \ G \int_{-\infty}^{s} \ ds' \build
\sum_{\ell \geq 2}^{} \varepsilon_{\ell} (s',n)$, so that we can directly use
the results of \cite{BD92} to relate the multipoles of the gravitational wave
at infinity to the multipoles of order $\ell \geq 2$ in $\varepsilon (s,n)$.}.

\section{Conclusion}

After having shown the mathematical and physical $(\forall x \in \Rb^4 , \
\partial_{\nu} \ T^{\mu \nu} (x)=0)$ consistency of the definition of massive
point-like photon rockets in special relativity, we have studied the retarded
linearized gravitational field that they generate. The crucial point is that
the source of the gravitational ra\-diation is the sum of the stress-energy
tensor of the matter and of that of the outgoing flux of photons. In fact, the
basic source of time-dependence in the system is the photon flux function
$\varepsilon (s,n)$. The acceleration of the matter is a derived quantity,
computable from the monopolar and dipolar pieces in the spherical harmonics
expansion of $\varepsilon (s,n)$: equations (3.13). We have proven by an
explicit calculation of the retarded potential generated by the total
stress-energy tensor $T^{\mu \nu} = T_{(m)}^{\mu \nu} + T_{(p)}^{\mu \nu}$ that
the amount of gravitational radiation at infinity is driven by the anisotropies
of order $\ell \geq 2$ in the photon flux $\varepsilon (s,n)$. On the other
hand, the coherent addition of the gravitational waves emitted by the monopolar
and dipolar pieces of the photon energy-tensor, and of the gravitational waves
emitted by the accelerated (because of the recoil) matter, exhibits a
cancellation leaving a time-dependent, but non-radiative gravitational field
(4.30). This cancellation is not too surprising if we consider, as recalled
above, that the photon flux function $\varepsilon (s,n)$ is the basic source of
time-dependence, and therefore gravitational radiation, in the system. In view
of the spin 2 of the graviton, it makes sense that only the anisotropies of
order $\ell \geq 2$ in $\varepsilon (s,n)$ can radiate gravitational waves.

\medskip

We have presented the calculation of the emission of gravitational radiation in
gory details to make it clear that there is absolutely no incompatibility
between the non-radiative character of the gravitational field generated by a
purely dipolar photon flux anisotropy and the perturbation techniques currently
used to compute the generation of gravitational radiation in general
relativity. Some of the technicalities of our treatment (notably the use of
\break analytic continuation techniques) have been forced upon us by the
necessity to deal in a mathematically rigourous way with the retarded
potentials
gene\-rated by peculiar singular sources $\propto \ \rho_R^{-2}$ (which become
infinite on a worldline and fall off rather slowly at infinity). A careless
treatment could miss the delicate compensation between the gravitational
radiation emitted by the Dirac-distributed matter energy tensor, and that
emitted by some of the singular terms $\propto \rho^{-3}$ appearing as
effective sources in the gauge-transformed photon-generated metric $h_{\mu
\nu}^{(p)} + \partial_{\mu} \ \xi_{\nu} +\partial_{\nu} \ \xi_{\mu}$.

\bigskip

\noindent {\bf Appendix}

\smallskip

\noindent To make contact with the literature let us remark that, as a by
product of the calculations above, we have shown that the linearized metric
$$
g_{\mu \nu} (x) = \eta_{\mu \nu} + h_{\mu \nu}^{'K} = \eta_{\mu \nu} + 2 \ G \
\frac{m(s_R)}{\rho_R} \ k_{\mu} \ k_{\nu} \ , \eqno ({\rm A}.1)
$$
satisfies everywhere in $\Rb^4$ (in the sense of distribution theory) the
linearized Einstein equations
$$
R_{\mu \nu}^{\rm lin} - \frac{1}{2} \ R^{\rm lin} \ \eta_{\mu \nu} = 8\pi
G \ \eta_{\nu \lambda} \ T_{\mu}^{\lambda} \ , \eqno ({\rm A}.2)
$$
with $T_{\mu}^{\nu}$ given by equations (3.1)-(3.3).

\medskip

Remarking now that (A.1) is of the Kerr-Schild form \cite{KS65}, and
remembering that G\"urses and G\"ursey \cite{GG75} have shown that all the
nonlinear terms cancel when Kerr-Schild metrics are inserted into the Einstein
equations written in terms of a {\it mixed} energy-momentum tensor
$T_{\mu}^{\nu}$, we see that we have rederived (in a roundabout way) the result
of Kinnersley \cite{K69}, namely that the metric (A.1) is, outside the
worldline $\Lc$, an exact solution of Einstein equations with source
$T_{(p)\mu}^{\nu} = (4\pi)^{-1} \ \rho_R^{-2} \ \varepsilon^K (s_R ,n) \
k_{\mu} \ k^{\nu}$. Note that, in the exact, nonlinear framework of Einstein's
equations, it makes no sense to write down $\delta$-like source terms, and
therefore to say that the exact Kinnersley metric describes a point-like
photon rocket. This is why, in the text, we consistently considered only
linearized gravity.

\end{document}